\documentclass[letterpaper,twocolumn,10pt]{article}
\usepackage{usenix-2020-09}

\usepackage{tikz}
\usepackage{amsmath}

\usepackage{filecontents}
\usepackage{cite}
\usepackage{amsmath,amssymb,amsfonts}
\usepackage{graphicx}
\usepackage{textcomp}
\usepackage{xcolor,colortbl}
\usepackage{fancyhdr}
\usepackage[noend]{algcompatible}
\usepackage{algorithm}
\usepackage{multirow}
\definecolor{Gray}{gray}{0.75}
\definecolor{LightGray}{gray}{0.85}
\definecolor{LightLightGray}{gray}{0.95}

\graphicspath{{images/}}
\DeclareGraphicsExtensions{.pdf,.jpeg,.png}

\begin{document}

\date{}

\title{\Large \bf Throughput Maximization of DNN Inference: Batching or Multi-Tenancy?}

\author{
{\rm Seyed Morteza Nabavinejad}\\
Worcester Polytechnic Institute\\
Worcester, MA\\
\and
{\rm Masoumeh Ebrahimi}\\
KTH Royal Institute of Technology\\
Sweden, Stockholm\\
\and
{\rm Sherief Reda}\\
Brown University\\
Providence, RI
} \maketitle
\begin{abstract}
Deployment of real-time ML services on warehouse-scale infrastructures is on the increase. Therefore, decreasing latency and increasing throughput of deep neural network (DNN) inference applications that empower those services have attracted attention from both academia and industry.   
  A common solution to address this challenge is leveraging hardware accelerators such as GPUs. To improve the inference throughput of DNNs deployed on GPU accelerators, two common approaches are employed: Batching and Multi-Tenancy. Our preliminary experiments show that the effect of these approaches on the throughput depends on the DNN architecture. Taking this observation into account, we design and implement \textit{DNNScaler} which aims to maximize the throughput of interactive AI-powered services while meeting their latency requirements. \textit{DNNScaler} first detects the suitable approach (Batching or Multi-Tenancy) that would be most beneficial for a DNN regarding throughput improvement. Then, it adjusts the control knob of the detected approach (batch size for Batching and number of co-located instances for Multi-Tenancy) to maintain the latency while increasing the throughput. Conducting an extensive set of experiments using well-known DNNs from a variety of domains, several popular datasets, and a cutting-edge GPU, the results indicate that \textit{DNNScaler} can improve the throughput by up to 14x (218\% on average) compared with the previously proposed approach, while meeting the latency requirements of the services.

\end{abstract}

\section{Introduction}
Deployment of interactive AI-powered services, also known as real-time ML, on warehouse-scale infrastructures is on the increase. The deep neural network (DNN) inference applications that empower these services have to meet the low-latency requirement of such real-time ML services. On the other hand, the service providers seek high throughput to serve more requests in a unit of time. They also desire high resource utilization to reduce their operational costs, and further improve their revenue. To this end, various hardware accelerators such as ASICs \cite{jouppi2017datacenter}, FPGA-based accelerators \cite{fowers2018configurable} and GPU-based accelerators \cite{hauswald2015djinn} are proposed for DNN inference. Since the GPUs have shown significant throughput improvement when employed for DNN inference, they are widely used in warehouse-scale infrastructures as DNN accelerators. \par

To gain high throughput when accelerating DNN inference, a common approach is Batching, which is widely used in previous works \cite{fang2017qos, shen2017escher}. It means processing input data in the form of batches, instead of processing them one by one. Batching helps to reuse the parameters of the DNN model for several inputs and also reduce the overhead of copying input data to GPU memory. \cite{shen2017escher, crankshaw2017clipper}. Another popular alternative is Multi-Tenancy \cite{jain2018dynamic, chen2016baymax}, where several different DNNs are co-located on a single GPU. Multi-Tenancy improves the throughput by sharing the computing resources between co-located DNNs. Although previous works have used Multi-Tenancy, they have not explored the case of co-locating several instances of the same DNN, in contrast to instances of different DNNs. In this work we consider this new approach for the first time (instances of the same DNN). Both Batching and Multi-Tenancy improve throughput via increasing resource utilization. While these approaches can increase the throughput, they negatively affect the tail latency of inference requests and elongate them \cite{jain2018dynamic}. Therefore they should be carefully used for real-time ML services.

In this paper, for the first time, we show the fact that the impact of Batching and Multi-Tenancy on the throughput depends on the DNN architecture. Based on the various features of a DNN, such as the number of parameters and computational complexity, either Batching or Multi-Tenancy can significantly improve the throughput of that DNN, while the other approach has no or negligible impact. Considering this observation, we design and implement our approach, called \textit{DNNScaler}, which aims to maximize the throughput of real-time ML services deployed on GPU accelerators while meeting their latency requirements. \textit{DNNScaler} consists of two modules: Profiler and Scaler. With the help of the Profiler module, it identifies the approach (Batching or Multi-Tenancy) that would be most beneficial for a DNN. After that, it adjusts the batch size (if Batching is selected) or the number of co-located instances (if Multi-Tenancy is selected) dynamically, considering the latency constraint, to maximize the throughput. Experimental results, using several DNNs and datasets and a Tesla P40 GPU, show that \textit{DNNScaler} can improve the throughput by up to 14x compared to an approach that ignores the impact of Batching and Multi-Tenancy on the throughput of different DNNs. We make the following contributions in this paper:

\begin{itemize}

\item We study the effect of Batching and Multi-Tenancy on throughput when deploying DNNs on a GPU accelerator. We examine various DNNs with varying architectures and features. By analyzing the results, for the first time, we show that the effectiveness of Batching and Multi-Tenancy highly depends on the DNN architecture. For some DNNs, Batching can significantly increase the throughput, while for others Multi-Tenancy remarkably improves throughput. In addition, to improve the throughput of a single DNN application, we suggest to deploy several instances of the same DNN, which is different from previous approaches that co-locate various DNNs on the same GPU.

\item We design a Profiler module that determines, at real-time, whether a DNN's throughput would benefit from Batching or Multi-Tenancy. Another designed module, Scaler, aims to maximize the throughput while maintaining latency. In the presence of the Batching approach, it tunes the batch size as a control knob to achieve its goal. The other control knob, the number of co-located DNN instances, is used by Scaler when Multi-Tenancy is chosen for improving throughput. In the Scaler module, we use machine learning to estimate the latency of the DNN for different number of co-located instances.    

\item Combining the Profiler and the Scaler modules, we implement our \textit{DNNScaler} approach. The Profiler module detects the suitable approach (Batching or Multi-Tenancy), and the Scaler module adjusts the respective control knob (batch size or the number of co-located instances). Conducting an extensive set of experiments using a wide variety of DNNs with different datasets as inputs and leveraging a powerful server equipped with an Nvidia GPU, we show the superiority of \textit{DNNScaler} over other approaches. 
\end{itemize} 

The rest of the paper is organized as follows: In Section \ref{sec:DNNInference}, we discuss the impact of Batching and Multi-Tenancy approaches on the throughput of various DNNs. Then, we introduce our proposed approach, \textit{DNNScaler}, in Section \ref{sec:method} and present the experimental results in Section \ref{sec:eval}. Related works are briefly discussed in Section \ref{sec:rela}, and the paper is concluded in Section \ref{sec:conc}.

\section{DNN Inference: Batching or Multi-Tenancy?}\label{sec:DNNInference}

The computing power and memory capacity of cutting-edge GPU accelerators used for DNN inference are on the increase. To improve the resource utilization of these accelerators, and hence, increase the throughput of applications, two common approaches are employed: \par

\textit{1) Batching:} In this approach, input data is processed in the form of batches instead of processing each individual input (e.g., each image in image classification DNNs) separately. This approach has been widely employed by previous works \cite{fang2017qos, song2017towards, tang2019nanily, crankshaw2017clipper} to increase the throughput by better utilizing the computing resources of GPUs. Since the weights of DNNs are needed at least once per each input, Batching helps to reuse them for multiple inputs and reduce the data copy to GPU memory \cite{song2017towards, shen2017escher}. 

\textit{2) Multi-Tenancy:} Since the DNN inference graphs used for prediction usually consume less resources than the available resources of GPUs, it is possible to deploy several instances of the same graph to potentially leverage instance-level parallelism and achieve higher resource utilization and throughput. Multi-Tenancy or co-location of several workloads or kernels on a single GPU and related challenges have been studied in a large body of research \cite{jain2018dynamic, chen2017effisha, zhang2019laius, chen2016baymax}. But none of them have considered multiple instances of the same DNN, as we consider in this work. 

We have conducted a set of experiments to understand the impact of these two approaches on throughput and latency of DNN inference. We have employed four image classification DNNs (described in Table \ref{tab:motiv}) with different sizes, architectures, and computational complexity to observe their performance under Batching and Multi-Tenancy. For the input data, we have used images from the ImageNet dataset \cite{russakovsky2015imagenet}. For obtaining the computational complexity of DNNs, we have used TensorFlow Profiler \cite{TensorFlowProfiler}. The GPU accelerator we have used is a Tesla P40 GPU that has 3840 CUDA cores and 24 GB GDDR5 memory. \par

For Batching, we use batch sizes of 1 to 128 to study its impact on throughput and latency. We have conducted the experiments for bigger batch sizes (up to 1024 that is supported by our GPU), but we only show the results for up to the batch size of 128 for the sake of clarity. For Multi-Tenancy, we co-locate 1 to 8 instances of the same DNN with increments of one (e.g., one instance of Inception-V1 to eight instances of it). For Multi-Tenancy, the batch size for all the instances is one.

\begin{table}[t]
\centering
\renewcommand{\arraystretch}{1}
\scriptsize
\caption{Number of Parameters and Computational Complexity of DNNs} 
\label{tab:motiv}
\begin{tabular}{lcc}
\hline
DNN & No. Parameters & \begin{tabular}[c]{@{}c@{}}Computational Complexity\\ of Inference (Mega FLOP)\end{tabular} \\ \hline
\rowcolor{LightLightGray} Inception-V1 & 6.6 M & 13.220736 \\
\rowcolor{LightGray} Inception-V4 & 42.7 M & 91.94925 \\
\rowcolor{LightLightGray} Mobilenet-V1-1 & 4.2 M & 8.420224 \\
\rowcolor{LightGray} ResNetV2-152 & 60.2 M & 120.084864 \\ \hline
\end{tabular}
\end{table}

\begin{figure}[t]
\centering
\includegraphics[width=\linewidth]{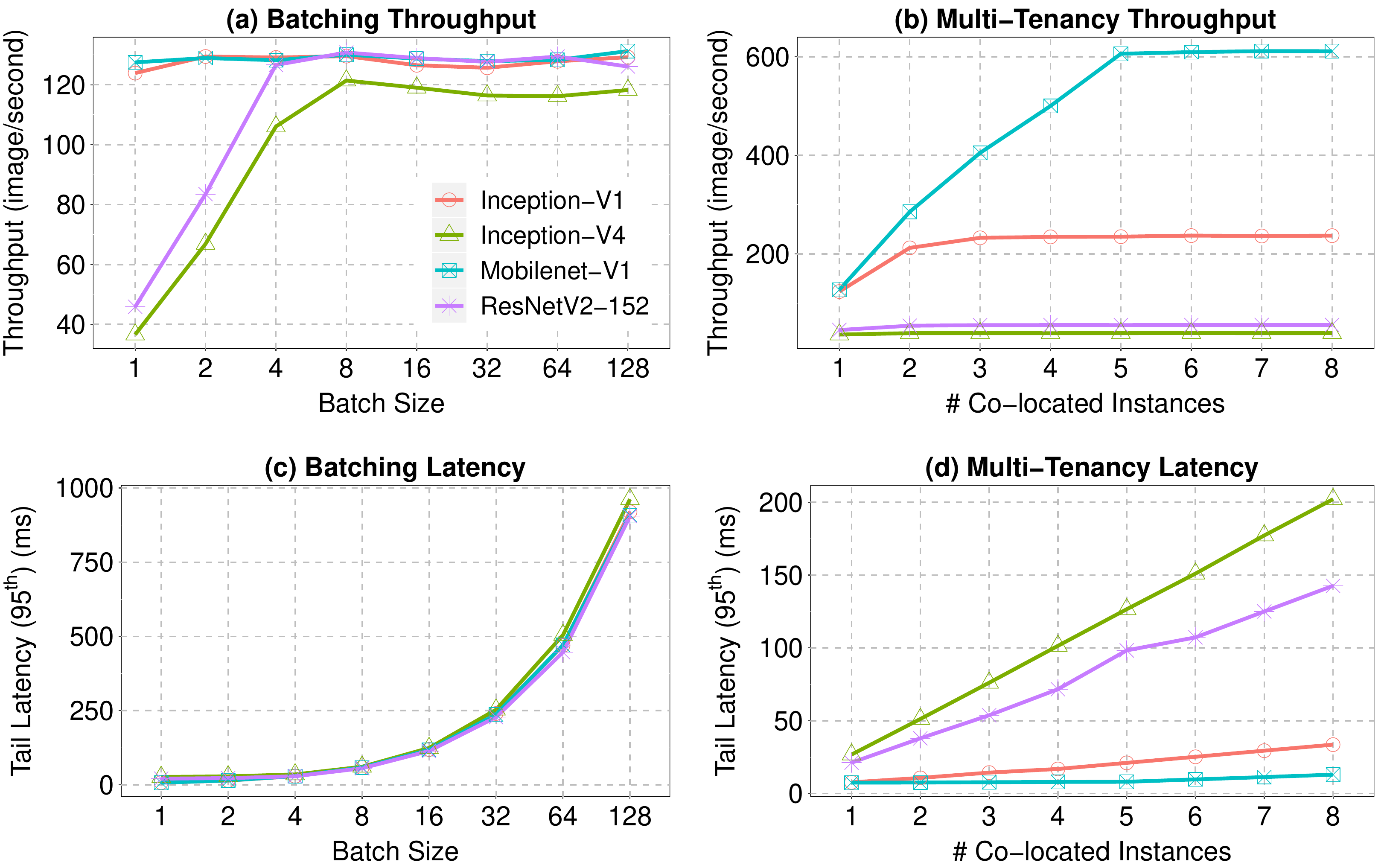}
\caption{Impact of Batching and Multi-Tenancy approaches on throughput and latency of different DNNs.} 
\label{fig:motiv1}
\end{figure}

The results are depicted in Fig. \ref{fig:motiv1}. As can be seen, different DNNs show different behavior under each approach. Batching (Fig. \ref{fig:motiv1}(a)) can significantly improve the throughput of Inception-V4 and ResNetV2-152. However, it has a negligible effect on the other two ones. On the other hand, Multi-Tenancy  (Fig. \ref{fig:motiv1}(b)) can improve the throughput of Inception-V1 and Mobilenet-V1-1, which could not leverage Batching. But, Multi-Tenancy has almost no effect on the throughput of Inception-V4 and  ResNetV2-152. We can also see the effect of Batching and Multi-Tenancy on the latency in Fig. \ref{fig:motiv1}(c) and (d), respectively. Tail latency is defined as the $95^{th}$ percentile of the inference latency distribution in this work. Both bigger batch size and a larger number of co-located instances lead to higher latency. Since latency is an essential requirement of real-time applications, it should be factored in when designing and implementing any approach. Combining the results presented in Fig. \ref{fig:motiv1} and the specifications of DNNs shown in Table \ref{tab:motiv} we conclude that: \par

1) Batching can significantly enhance the throughput of DNNs with a large number of parameters and high computational complexity (e.g., Inception-V4). In these networks, Batching helps to reuse the parameters for several inputs, and hence, reduce the data movement in GPU, which leads to an increase in throughput. For DNNs with a small number of parameters such as Inception-V1, however, Batching is not very effective. In these DNNs, the time needed for preparing and copying the input data to GPU dominates the time needed for copying the parameters, and hence, parameter reuse cannot improve the throughput noticeably. To observe this, we profile the share of kernels launched during execution of DNNs for Inception-V1 and Inception-V4. Results show that share of kernels related to data preparation and movement (e.g., \texttt{redzone-checker, CUDA memcopy HtoD}) in the total execution time is very significant in Inception-V1 (20.1\% only for aforementioned kernels for BS = 16), and it becomes even more when increasing the batch size. For Inception-V4, however, those kernels do not consume a large portion of total execution time (e.g., 4.2\% for BS = 16). For the profiling, we used the NVProf tool \cite{nvprof}. Due to large number of kernels profiled (68) and lack of space, we only presented the info for two kernels.

\begin{figure}[t]
\centering
\includegraphics[width=.9\linewidth]{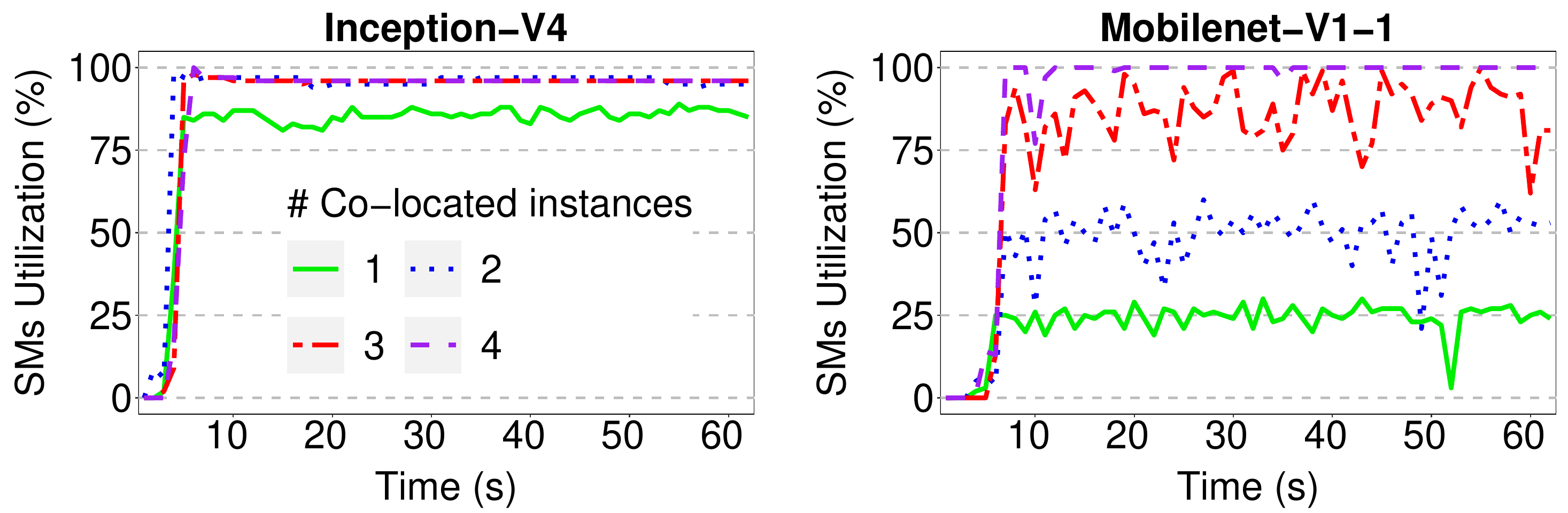}
\caption{Impact of co-location on streaming multiprocessors (SMs) utilization for two DNNs.}\label{fig:motiv2}
\end{figure}

2) Small and simple DNNs with low computation requirements can most benefit from Multi-Tenancy. Since the GPU accelerators have large amount of computing resources, one instance of small DNNs cannot fully utilize them. Hence, the idle resources can be used by additional instances to process several inputs simultaneously. Therefore these DNNs can experience significant throughput improvement by Multi-Tenancy. The complex networks such as Inception-V4, however, utilize all or most of the computing resources of GPU by only one instance. Hence, co-locating several instances of them cannot yield throughput improvement since the instances should utilize the computing resources in a time-sharing manner. The resource utilization of GPU under the Multi-Tenancy approach (from one to four co-located DNN instances) for Mobilenet-V1-1 and Inception-V4 is depicted in Fig. \ref{fig:motiv2}. 

Considering the aforementioned conclusions derived from preliminary experiments, we design and implement our approach which is discussed in detail in the next section.

\section{Methodology}\label{sec:method}

Our approach, \textit{DNNScaler}, aims to maximize the throughput while meeting the latency constraint of real-time ML applications, by leveraging either Batching or Multi-tenancy based on the target DNN. First, we present the problem formulation and then describe the design and implementation of \textit{DNNScaler} in detail. The acronyms used in the paper and their meaning are listed in Table \ref{tab:acro}.

\begin{table}[t]
\centering
\caption{Lists of Acronyms and Their Meaning Used Throughout the Paper} 
\label{tab:acro}
\scriptsize
\begin{tabular}{ll}
\hline
Acronym & Definition \\ \hline
B & Batching \\
BS & Batch Size \\
MT & Multi-Tenancy \\
MTL & Multi-Tenancy Level \\
DNN & Deep Neural Network \\
TI & Throughput Improvement \\
SLO & Service Level Objective \\
SM & Streaming Multiprocessor \\
FLOP & Floating Point Operation \\ \hline
\end{tabular}
\end{table}

\subsection{Problem Statement and Formulation}\label{subsec:prob}

A DNN inference application is deployed on a GPU accelerator with a latency constraint stated in the form of Service Level Objective \textit{(SLO)}. 
Both throughput and latency of DNN are a function of the Batch Size (BS) or the Multi-Tenancy Level (MTL), 
depending on which approach is chosen (Throughput $\propto f(BS, MTL)$, Latency $\propto g(BS, MTL)$). By MTL, we mean the number of co-located instances of the DNN on the GPU.
The objective function is to maximize the throughput of the DNN during its execution time $T$, while maintaining its latency below the SLO. 

\footnotesize
\begin{equation}\label{eq1}
Maximize\quad \dfrac{1}{T}\sum_{t = 1}^{T}Throughput^{t}
\end{equation}
s.t.
\begin{equation} \label{eq2}
Latency^{t} \leq SLO
\end{equation}
\normalsize

This problem provides two control knobs for managing the throughput and latency: 1) We can use either Batching or Multi-Tenancy for increasing the throughput. 2) Depending on which one is selected, we fine-tune the batch size (for Batching) or the number of co-located instances (for Multi-Tenancy) to maintain the latency.

\subsection{DNNScaler}\label{subsec:DNNScaler}

\begin{figure*}[t]
\centering
\includegraphics[width=.9\linewidth]{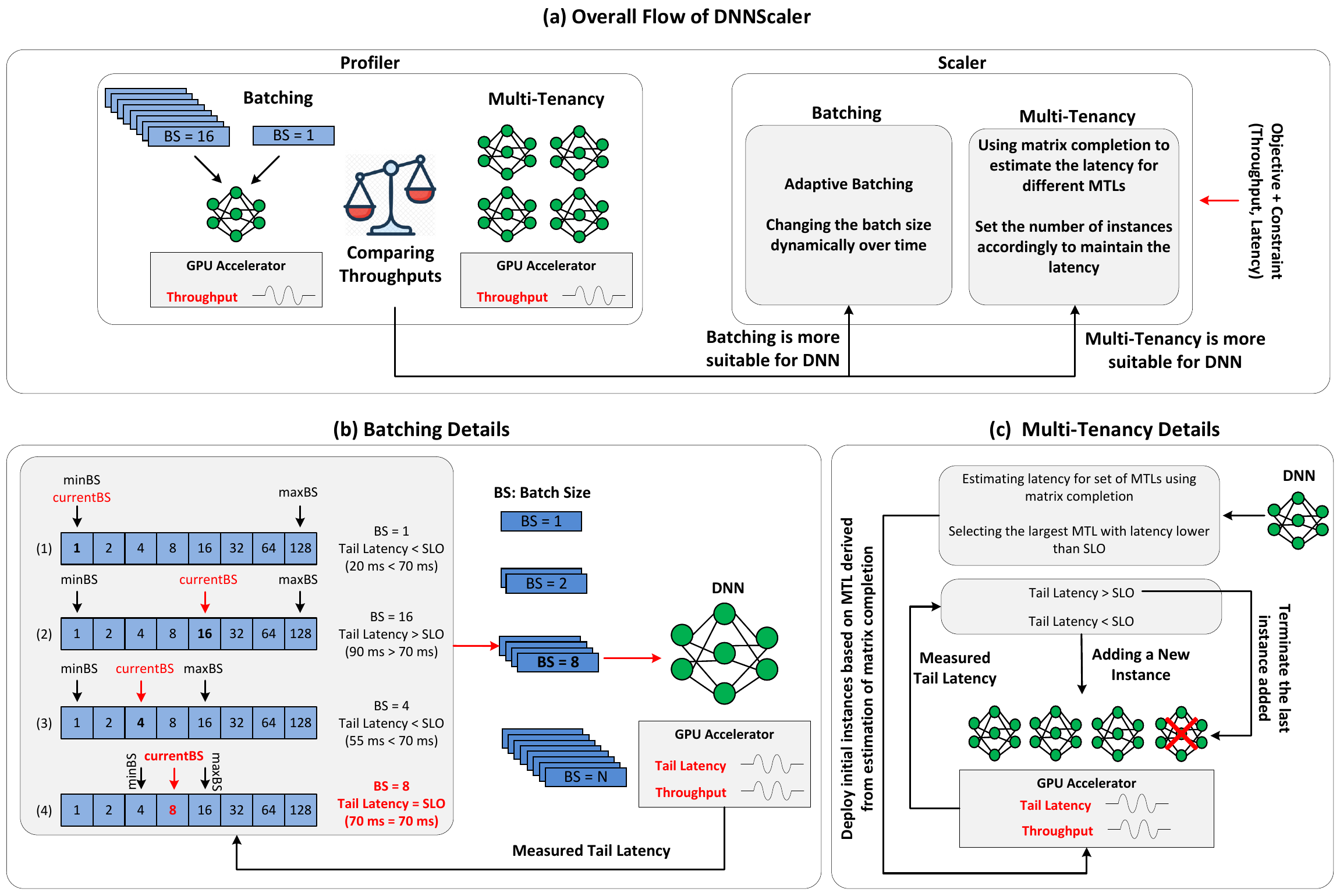}
\caption{The overall flow of DNNScaler and its Profiler and Scaler modules are depicted in (a). The detail flow of The Scaler for Batching and Multi-Tenancy approaches is depicted in (b) and (c), respectively.} \label{fig:flow}
\end{figure*}

Our proposed approach, \textit{DNNScaler}, leverages the observations discussed in Section \ref{sec:DNNInference} when using the two control knobs (Batching and the batch size, or Multi-Tenancy and the number of co-located instances). Since Batching and Multi-Tenancy in this context sparks a sense of scaling-up and scaling-out of DNN inference applications, respectively, we have chosen the name \textit{DNNScaler} for our approach. \textit{DNNScaler} consists of two modules: Profiler and Scaler. The overall flow of \textit{DNNScaler} is shown in Fig. \ref{fig:flow}(a), and its pseudo-code is presented in Algorithm 1. In the following, we explain these two modules.

\subsubsection{\textbf{Profiler}}
The Profiler module probes the DNN to determine which of Batching or Multi-Tenancy can better improve the throughput. To determine which approach is more suitable for a DNN, the Profiler conducts a lightweight profiling at run-time. During this profiling phase, the throughput of the DNN for batch sizes of one (BS = 1) and $m$ (where $m = 32$ in our experiments) is measured by the Profiler. Only a few batches are needed to be executed to measure the throughput of each BS, and calculate the throughput improvement obtained by $BS = m$ over $BS = 1$, as in (\ref{eq3}):

\footnotesize
\begin{equation}\label{eq3}
TI_{B} = \frac{Throughput_{BS=m} - Throughput_{BS=1}}{Throughput_{BS=1}} \times 100 
\end{equation}
\normalsize

After that, the throughput for the case of having $n$ co-located instances (MTL = n, where $n = 8$ in our experiments) is measured. The throughput for a single instance is not needed because it is the same as Batching with BS = 1. Then, throughput improvement of Multi-Tenancy can be calculated as (\ref{eq4}). Comparing the throughput improvement of Batching and Multi-Tenancy (see (\ref{eq5})) , the Profiler decides which approach is more suitable for the DNN and sends the gathered information to the next module, Scaler. The profiling is of the order of seconds, therefore its overhead on the system is negligible. 

\footnotesize
\begin{equation}\label{eq4}
TI_{MT} = \frac{Throughput_{MTL=n} - Throughput_{MTL=1}}{Throughput_{MTL=1}} \times 100
\end{equation} 

\begin{equation}\label{eq5}
if 
\begin{cases}
    TI_{B} > TI_{MT} , & \text{Batching} \\
    TI_{B} < TI_{MT},              & \text{Multi-Tenancy}\\
    TI_{B} = TI_{MT} , & \text{The one with lower latency} 
\end{cases}
\end{equation}
\normalsize

\subsubsection{\textbf{Scaler}}\label{subsubsec:scaler}
The Scaler module receives the information from the Profiler that indicates which approach (Batching or Multi-Tenancy) is appropriate. Having this information, the Scaler aims to maintain the SLO of the DNN while trying to maximize its throughput. Looking again at Fig. \ref{fig:motiv1}, we see that for both Batching and Multi-Tenancy, increasing the batch size and the number of co-located instances can lead to higher throughput, but simultaneously leads to elongated latency. Hence, the Scaler module tries to find the largest batch size or the number of co-located instances (based on the approach proposed by the Profiler) that yields the latency below or equal to SLO. In the following, we describe how the Scaler module works with respect to the selected approach.

\subsection{Dynamic Behavior of Scaler}
In this section, we describe the dynamic behavior of the Scaler module of \textit{DNNScaler} with respect to the approach determined by the Profiler module for a job. First, we discuss how Scaler adjusts the batch size when the Batching approach is selected. Then, we describe the Scaler mechanism for the case when Multi-Tenancy is selected for a job and explain how the number of co-located instances is determined dynamically with respect to latency and throughput. 
\subsubsection{Dynamic Batch Size Adjustment} When deploying a DNN on the GPU for inference, the common practice is to use a constant batch size. This constant batch size cannot be changed during the execution dynamically. In order to change it, the current instance should be terminated and a new one with another batch size should be launched, which imposes overhead on the system in the form of interruption in the service, increased latency, and reduced throughput. To address this issue, we implement dynamic batch sizing for DNN inference. The implementation imposes almost no overhead on latency or throughput, compared with a conventional constant batch size approach. Implementing the dynamic batch sizing helps us to design and implement the Scaler module more efficiently. The changes are in application side and how it interacts with the TensorFlow framework, without any need for changing TensorFlow.\par

In the design of the Scaler for the Batching approach, we consider the observation presented in Section \ref{sec:DNNInference}. 
We saw that both latency and throughput have a direct relationship with batch size. The Scaler leverages this observation and employs a pseudo binary search mechanism to efficiently search for the most suitable batch size.
The time complexity of the binary search is \textit {O (log n)}, and hence, the time overhead of Scaler would be negligible.

As shown in Fig. \ref{fig:flow}(b), the Scaler module for the Batching approach works as follows: it starts with a default batch size of one (BS=1) and processes a certain number of batches and measures their tail latency. If the tail latency is less than the SLO of the DNN multiplied by a $\alpha$ coefficient ($SLO \times \alpha$), then Scaler sets the batch size equal to the value in the median of the current batch size and the largest possible batch size (BS=128 in this work). If the current batch size is the largest possible one (due to the limitation of GPU memory, the batch size cannot be larger than a certain value), then it means no further throughput improvement is possible. 
Otherwise, if the tail latency is greater than the SLO, the Scaler sets the batch size as the value in the median of the lowest possible batch size and the current batch size.  In this case, the current batch size being the smallest one means that the SLO of the DNN cannot be met.
Finally, if the latency is between SLO and $SLO \times \alpha$, then Scaler does not change anything and continues with the current batch size. We use $\alpha$ coefficient to avoid excessive batch size changes. The suitable value of $\alpha$ can be found empirically by observing the behavior of a few DNNs under different values of it. We have $\alpha = 0.85$ in this work.\par

The Scaler does not stop after finding a suitable batch size, but continues to monitor the latency. Detecting a tail latency above the SLO or below $SLO \times \alpha$, it starts adjusting the batch size again. Readjustment is needed when the tail latency is affected by parameters such as variation in the input dataset, GPU temperature, and frequency of GPU. Even the user can decide to change the SLO during runtime. 

\subsubsection{Dynamic MT Level Adjustment}
Multi-tenancy shows a similar behavior to batching (for the DNNs that can benefit from it). Increasing the number of co-located instances can improve the throughput (and also increases the latency). Similar to batching that we used BS = 128 as the upper bound of batch size, for multi-tenancy we have chosen MTL = 10 as the maximum number of co-located instances based on the memory capacity of our GPU. This number also can be determined for various settings (such as different GPUs) by a lightweight profiling. Therefore a similar approach to Batching (binary search) can be employed to find the best value for MTL (i.e., the number of co-located instances). However, unlike Batching where we implemented a dynamic batch sizing with negligible overhead, for Multi-Tenancy there is no such lightweight mechanism to change the number of co-located instances on the fly. \par

Frequently launching and terminating instances imposes significant overhead. Therefore, we need an alternative approach with low overhead. One solution is to profile the latency of the DNN for all the possible number of instances (MTL = 1 to MTL = N). In the next step, to adjust the value of MTL for a specific SLO, we can simply select the largest one (to maximize throughput) that has latency lower than SLO. However, the overhead of profiling all the possible values of MTL itself leads to significant overhead, which is in contrast with our initial goal, which was to avoid overhead of frequent launching and terminating instances.\par

\begin{figure*}
\centering
\includegraphics[width=.9\linewidth]{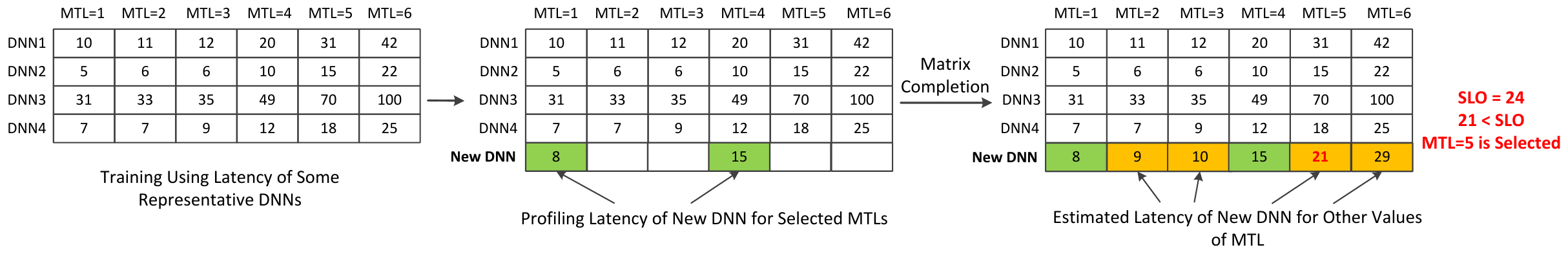}
\vspace{-10pt}
\caption{Illustrative example to show how we employ matrix completion to estimate the latency of a DNN for different MTLs.}\label{fig:matrix}
\vspace{-5pt}
\end{figure*}

To tackle this challenge, we employ a machine learning based approach called matrix completion \cite{candes2009exact} to estimate the latency for all the possible values of MTL. Using matrix completion, we need to profile results of the latency of DNN for a few values of MTL (two in our work). Since we already have this information from the profiling phase (for MTL = 1 and MTL = 8), we do not impose any further overhead to the system. Having the latency of MTL = 1 and MTL = 8, matrix completion can estimate the latency for other number of co-located instances (i.e., other values of MTL). Then, we use these estimated values to select the MTL considering the SLO. Fig. \ref{fig:matrix} shows how the matrix completion is employed to estimate the latency of different MTLs. With matrix completion we can jump to a solution immediately without changing the value of MTL frequently, in contrast to a brute force approach.\par

Since the estimated values of matrix completion are not 100\% accurate, we have devised an additive-increase-multiplicative-decrease (AIMD) scheme\cite{crankshaw2017clipper, chiu1989analysis} to complement it. We start the co-location with MTL suggested by matrix completion. If the latency is lower than SLO, then we start adding instances one by one until tail latency is greater than SLO. In this point, we only need to terminate the last instance to keep the tail latency below the constraint. That is the point where we can have the highest possible throughput while maintaining the SLO. If we reach to the maximum MTL, e.g., MTL = 10, (where no further instances can be deployed) before violating the SLO, we can stop adding new instances and there would be no need for terminating any instance. On the other hand, if the latency of MTL suggested by matrtix completion is higher than SLO (which means that latency is underestimated), we decrease the number of instances by steps of one and terminate them, until the latency is lower than SLO, and then stop. The flow of Scaler for Multi-tenancy approach is presented in Fig. \ref{fig:flow}(c).

Note that the main difference between our work and previous Multi-Tenancy approaches, which forms one of the contributions of the paper, is that they consider DNNs co-located from different jobs and try to mitigate the impact of interference on their performance. But we consider the case where the co-located instances are from the same DNN and belong to the same job and work with each other to improve the total throughput. 

For the BS, we used binary search to find upper bound of it (128)  that does not lead to out of memory (OOM) error by a few short experiments. For finding upper bound of MTL (10), first the minimum amount of memory needed for a DNN is determined considering its size and computational complexity, and then the upper value of MTL is calculated based on this value (for the largest DNN) and the memory capacity of GPU (and also overhead of several instances working together).

\textbf{Matrix Completion}, an ML approach, is used to recover missing entries
of a matrix that is partially observed. It employs Singular Value Decomposition (SVD) to reduce the dimensions of the matrix. It also needs to know the rank of the matrix of interest. The rows or columns of matrix with rank $r$ span an $r$-dimensional space. Applying SVD on the matrix $M$ yields a factorization of the form $M = \boldsymbol{U}\times \boldsymbol{\Sigma} \times \boldsymbol{V}^{\boldsymbol{T}}$, where~$\boldsymbol{U}$ ,~$\boldsymbol{V}$ and~$\boldsymbol{\Sigma}$ represent different similarity concepts features of $M$.  

\scriptsize
\begin{align}\nonumber
\boldsymbol{U}_{n_1\times r}=
\begin{bmatrix}
    \boldsymbol{u}_{11} &   \dots  & \boldsymbol{u}_{1r} \\
    \boldsymbol{u}_{21} &   \dots  & \boldsymbol{u}_{2r} \\
    \vdots &   \ddots & \vdots \\
    \boldsymbol{u}_{n_11} &   \dots  & \boldsymbol{u}_{n_1r}
\end{bmatrix},
\boldsymbol{V}_{n_2 \times r}=
\begin{bmatrix}
    \boldsymbol{v}_{11} &   \dots  & \boldsymbol{v}_{1r} \\
    \boldsymbol{v}_{21} &  \dots  & \boldsymbol{v}_{2r} \\
    \vdots &  \ddots & \vdots \\
    \boldsymbol{v}_{n_21} &  \dots  & \boldsymbol{v}_{n_2r}
\end{bmatrix},
\end{align}
and
\begin{align}\nonumber
\boldsymbol{\Sigma}_{r \times r}=
\begin{bmatrix}
    \boldsymbol{\sigma}_{1}  & \dots  & 0 \\
    \vdots & \ddots & \vdots \\
    0 & \dots  & \boldsymbol{\sigma}_{r}
\end{bmatrix}.
\end{align}
\normalsize

Having matrices~$\boldsymbol{U}$,~$\boldsymbol{\Sigma}$, and~$\boldsymbol{V}$ and applying PQ-reconstruction leads to matrix~$X$ that estimates the missing values of $M$. We use convex optimization by  TFOCS (Templates for First-Order Conic Solvers)~\cite{becker2011templates} tool to estimate matrix~$X$ in this work.

\begin{algorithm}[t]
\caption{DNNScaler}
\label{alg:dnnscaler}
\begin{algorithmic}[1]
\scriptsize
\STATEx \textbf{Input: }SLO, DNN Model
\Statex
\STATEx \textbf{Profiler}
\STATE $TI_{B}$ = $Throughput_{BS=m} - Throughput_{BS=1}$ 
\STATE $TI_{MT} = Throughput_{MTL=n} - Throughput_{MTL=1}$ 
\IF {$TI_{B} > TI_{MT}$}
\STATE Batching is selected for the DNN
\ELSIF  {$TI_{B} < TI_{MT}$}
\STATE Multi-Tenancy is selected for the DNN
\ELSIF {$TI_{B} = TI_{MT}$}
\STATE The one with lowest latency is selected for the DNN
\ENDIF 
\Statex
\STATEx \textbf{Scaler}
\IF {Batching is selected}
\STATE minBS, initialBS, currentBS = 1, maxBS = N, BS = SB(1), LatencyList = []
 \WHILE{True}
 \STATE LatencyList.append(monitorLatency(inference(BS)))
 \IF{$\alpha \times SLO \leq$ max(LatencyList) $\leq SLO$ }
 \STATE Continue with BS
 \ELSIF{ max(LatencyList) $< \alpha \times SLO$ }
 \STATE minBS = currentBS
 \STATE currentBS = ceil($\frac{minBS+maxBS}{2}$)
 \STATE BS = SB(currentBS)
 
 \ELSIF{max(LatencyList) $> SLO$}
 \IF{currentBS = 1}
 \STATE Further BS reduction is not possible.
 \ENDIF
 \IF{currentBS = minBS}
 \STATE maxBS= currentBS, minBS = 1, 
 \STATE currentBS = floor($\frac{minBS+maxBS}{2}$)
 \STATE BS = SB(currentBS)
 \ELSE
 \STATE maxBS = currentBS
 \STATE currentBS = floor($\frac{minBS+maxBS}{2}$)
 \STATE BS = SB(currentBS)
 \ENDIF
 \ENDIF
 \ENDWHILE
\ELSIF {Multi-Tenancy is selected}
\STATE Estimate DNN Latency using Matrix Completion
\STATE MTL = Max $\{1..N\}$, where Latency $<$ SLO
\STATE LatencyList.append(monitorLatency(inference(MTL)))
\IF{$\alpha \times SLO \leq$ max(LatencyList) $\leq SLO$ }
\STATE Continue with MTL
\ELSIF { max(LatencyList) $< \alpha \times SLO$ }
\STATE MTL = MTL + 1
\STATE  Launch a new DNN instance on GPU
 \ELSIF{max(LatencyList) $> SLO$}
 \STATE MTL = MTL - 1
 \STATE Terminate the last added DNN instance
\ENDIF
\ENDIF
\end{algorithmic}
\end{algorithm}

Since \textit{DNNScaler} is proposed for real-time applications, it is very important that they experience little to no interrupt in their work. Both the Batching and Multi-Tenancy approaches of \textit{DNNScaler} assure that the applications can continue their work with no interruption. None of adjusting the batch size or the number of co-located instances prevents the DNNs from serving the new requests. Moreover, both mechanisms can quickly respond to bursty workloads and avoid violating latency constraints, as some inference workloads arrive in a burst and not uniformly \cite{ali2020batch, DigitalInsight}.

\section {Evaluation}\label{sec:eval}

\subsection {Experimental Setup} \label{subsec:exp}

\noindent \textbf{Platform.} We run our experiments on a dual-socket Xeon server. It has two  E5-2680 v4 Xeon chips where each of the chips has 28 cores running at 2.4 GHz. The server has 128 GB of DDR4 memory. Ubuntu 16.04 with kernel 4.4 is installed on the server with Python 2.7, CUDA 11.0, and TensorFlow 1.15. The server is equipped with a PCI Express Gen3 Nvidia Tesla P40 GPU Accelerator. The Tesla P40  leverages Nvidia Pascal architecture and has 3840 CUDA cores. The total memory capacity of GPU is  24 GB GDDR5 memory, its idle power is around 50W, and its maximum power limit is 250W. 

\noindent \textbf{Networks and Dataset}. To show the adaptive nature of our approach, we use DNNs from different domains. Since computer vision, and in particular image classification, is a popular field and numerous DNNs are designed for this application, we employ 16 image classification networks with two datasets in our experiments. One dataset is ImageNet \cite{russakovsky2015imagenet}, which is a popular dataset that is widely used in previous works \cite{hill2017deftnn,  kim2019mulayer, marco2020optimizing, dhakal2020gslice}, and the other one is Caltech-256 \cite{griffin2007caltech}, which is collected by researchers from the California Institute of Technology. For these DNNs, throughput is defined as number of images processed per second (image/second). From the natural language processing (NLP) domain, we employ a DNN for text classification \cite{DBLP:journals/corr/Kim14f}, which we call TextClassif in this paper. For the input data of this DNN, we use Sentiment140 \cite{sentiment} and IMDB Reviews \cite{maas-EtAl:2011:ACL-HLT2011} datasets. For this DNN, the throughput is defined as the number of sentences processed per second. DeepVS \cite{jiang2018deepvs} is another DNN we use in our experiments that targets video saliency prediction and the throughput is defined as the number of frames processed per second. Finally, we employ DeepSpeech2 \cite{amodei2016deep}, which is an end-to-end DNN for speech recognition and define the throughput as the number of speech files processed per second. The selected DNNs cover a wide range of applications, as well as DNN types: from CNNs to RNNs, to LSTMs. These DNNs have varying sizes and architectures, and consequently, different computational complexity. The specifications of the networks and datasets are presented in Table \ref{tab:DNNs}. 

\noindent \textbf{System Comparison.}
Clipper \cite{crankshaw2017clipper} is an approach proposed for online serving of inference requests considering a predefined latency SLO. Clipper employs an additive-increase-multiplicative-decrease (AIMD) scheme to find the optimal batch size that maximizes the throughput while meeting the latency SLO. It starts from the minimum batch size and additively increases it by a fixed step (four in this work) until tail latency surpasses the SLO. At this point, Clipper performs a small multiplicative back-off and reduces the BS by 10\%.
\begin{table}[]
\centering
\caption{Lists of DNNs Used in the Experiments} 
\label{tab:DNNs}
\scriptsize
\setlength{\tabcolsep}{3pt}
\begin{tabular}{lccc}
\hline
DNN & Abberviation & Domain & Dataset \\ \hline
Inception-V1 \cite{szegedy2015going} & Inc-V1 & \multirow{16}{*}{\rotatebox{90}{Image Classification}} & \multirow{16}{*}{\rotatebox{90}{ImageNet \& Caltech-256}} \\
Inception-V2 \cite{ioffe2015batch}  & Inc-V2 &  &  \\
Inception-V3 \cite{szegedy2016rethinking}  & Inc-V3 &  &  \\
Inception-V4 \cite{szegedy2017inception}  & Inc-V4 &  &  \\
Mobilenet-V1-1 \cite{howard2017mobilenets}  & MobV1-1 &  &  \\
Mobilenet-V1-05 \cite{howard2017mobilenets}  & MobV1-05 &  &  \\
Mobilenet-V1-025 \cite{howard2017mobilenets}  & MobV1-025 &  &  \\
Mobilenet-V2-1  \cite{sandler2018mobilenetv2} & MobV2-1 &  &  \\
Mobilenet-V2-14  \cite{sandler2018mobilenetv2} & MobV2-14 &  &  \\
NASNET-Large  \cite{zoph2018learning} & NAS-Large &  &  \\
NASNET-Mobile  \cite{zoph2018learning} & NAS-Mob &  &  \\
PNASNET-Large  \cite{liu2018progressive} & PNAS-Large &  &  \\
PNASNET-Mobile  \cite{liu2018progressive} & PNAS-Mob &  &  \\
ResNet-V2-50  \cite{he2016identity} & ResV2-50 &  &  \\
ResNet-V2-101  \cite{he2016identity} & ResV2-101 &  &  \\
ResNet-V2-152 \cite{he2016identity} & ResV2-152 &  &  \\ \hline
TextClassif \cite{DBLP:journals/corr/Kim14f} & - & NLP & \begin{tabular}{@{}c@{}}Sentiment140 \\ IMDB Review\end{tabular} \\ \hline
DeePVS \cite{jiang2018deepvs} & - & \begin{tabular}{@{}c@{}}Video \\ Saliency\end{tabular} & \begin{tabular}{@{}c@{}}LEDOV \cite{jiang2018deepvs} \\ DHF1K \cite{Wang_2018_CVPR, Wang_2019_revisitingVS}\end{tabular} \\ \hline
DeepSpeech2 \cite{amodei2016deep} & DeepSpeech & \begin{tabular}{@{}c@{}}Speech \\ Recognition\end{tabular}  & LibriSpeech \cite{panayotov2015librispeech} \\ \hline
\end{tabular}
\end{table}

\noindent \textbf{Workload.} In our experiments, we have a workload consists of 30 DNN inference jobs. SLO of each job is stated in the form of $95^{th}$ tail latency target in milliseconds. We measured the average latency of one input for BS = 1 and MTL = 1. Then, we considered a coefficient (> 1) of this value for SLO of each job to have tight and relaxed SLOs. The list of jobs is presented in Table \ref{tab:jobs}.

\begin{table*}[t]
\centering
\caption{Specification of jobs used in the experiments. The "DNNScaler Method" column is filled after applying our method. The last column ("Steady MTL/BS") shows the steady state batch size (BS) or number of multi-tenant instances (MTL) that DNNScaler has chosen for each job in the experiments.} 
\label{tab:jobs}
\renewcommand{\arraystretch}{1}
\setlength\tabcolsep{5pt} \scriptsize
\begin{tabular}{lccccc|lccccc}
\hline
Job \# & DNN       & Dataset  & SLO (ms) & \begin{tabular}[c]{@{}c@{}}DNNScaler\\ Method\end{tabular} & Steady MTL/BS & Job \# & DNN         & Dataset      & SLO (ms) & \begin{tabular}[c]{@{}c@{}}DNNScaler\\ Method\end{tabular} & Steady MTL/BS \\ \hline
1      & Inc-V1    & ImageNet & 35       & MT                                                         & MTL = 8       & 16     & Inc-V3      & CalTech      & 322      & B                                                          & BS = 37       \\
2      & Inc-V2    & ImageNet & 53       & MT                                                         & MTL = 9       & 17     & Inc-V4      & CalTech      & 139      & B                                                          & BS = 10       \\
3      & Inc-V4    & ImageNet & 419      & B                                                          & BS = 28       & 18     & MobV1-1     & CalTech      & 89       & MT                                                         & MTL = 10      \\
4      & MobV1-05  & ImageNet & 199      & MT                                                         & MTL = 10      & 19     & MobV1-05    & CalTech      & 60       & MT                                                         & MTL = 10      \\
5      & MobV1-025 & ImageNet & 186      & MT                                                         & MTL = 10      & 20     & MobV1-025   & CalTech      & 104      & MT                                                         & MTL = 10      \\
6      & MobV2-1   & ImageNet & 81       & MT                                                         & MTL = 10      & 21     & MobV2-1     & CalTech      & 129      & MT                                                         & MTL = 10      \\
7      & NAS-Large & ImageNet & 417      & B                                                          & BS = 13       & 22     & PNAS-Large  & CalTech      & 524      & B                                                          & BS =19        \\
8      & NAS-Mob   & ImageNet & 85       & MT                                                         & MTL = 10      & 23     & PNAS-Mob    & CalTech      & 321      & B                                                          & BS = 50       \\
9      & PNAS-Mob  & ImageNet & 82       & MT                                                         & MTL = 10      & 24     & ResV2-50    & CalTech      & 31       & B                                                          & BS = 1        \\
10     & ResV2-50  & ImageNet & 45       & MT                                                         & MTL = 6       & 25     & ResV2-101   & CalTech      & 107      & B                                                          & BS = 10       \\
11     & ResV2-101 & ImageNet & 72       & B                                                          & BS = 4        & 26     & TextClassif & Sentiment140 & 3.5      & B                                                          & BS = 102      \\
12     & ResV2-152 & ImageNet & 206      & B                                                          & BS =14        & 27     & TextClassif & IMDB         & 3        & B                                                          & BS = 76       \\
13     & ResV2-101 & ImageNet & 107      & B                                                          & BS = 7        & 28     & DeepSpeech  & LibriSpeech  & 1250     & B                                                          & BS = 28       \\
14     & Inc-V1    & CalTech  & 48       & MT                                                         & MTL = 10      & 29     & DeePVS      & LEDOV        & 3000     & MT                                                         & MTL = 6       \\
15     & Inc-V2    & CalTech  & 116      & B                                                          & BS = 16       & 30     & DeePVS      & DHF1K        & 5000     & MT                                                         & MTL = 8       \\ \hline
\end{tabular}
\end{table*}

\subsection{Profiling Results}\label{subsec:prof}

We have profiled the DNNs using the Profiler module to identify the proper approach for each of them. For Batching, we use BS=1 and BS=32, and for Multi-Tenancy we use MTL = 1 and MTL = 8. These values (BS = 32 and MTL= 8) are chosen based on our early observations. We have seen that BS = 32 and MTL = 8 are big enough to show which approach can give higher throughput improvement. These can be chosen differently for other GPUs or DNNs, if needed. The percentage of improvement (over base throughput of MTL = 1 and BS = 1) yield by each approach for several jobs is shown in Table \ref{tab:profIma}. The DNNScaler Method column in Table \ref{tab:jobs} is filled by the results obtained from profiling. The results further emphasize our observations from preliminary experiments that one of the Batching or Multi-Tenancy works better for a DNN in terms of throughput improvement. For networks with a low amount of computational complexity and a low number of parameters, such as the Mobilenet, we see remarkable throughput improvement (e.g., 335\% in Job 19) by Multi-Tenancy, while the same DNNs cannot benefit from Batching significantly. On the other hand, large and complex networks with a high number of parameters such as Inception-V4 can experience high throughput improvement by Batching, but not by Multi-Tenancy (see Job 3).\par 

As can be seen, the dataset also affects the performance of Batching and Multi-Tenancy, and hence, the approach selected for the DNNs. For example, in image classification DNNs, the image size is important as it should be readjusted before being fed to the network. This adjustment depends on the dataset, and affects the overall performance of DNN. Therefore, for some DNNs such as Inception-V2, the Multi-Tenancy approach yields better throughput for the ImageNet dataset, but Batching has better performance for the Caltech dataset. The length of sentences also affects the performance of TextClassif, and hence, it shows different behavior for Sentiment140 and IMDB Reviews datasets with respect to latency. The longer sentences of IMBD Reviews take more time to be processed.

\begin{table}[]
\centering
\caption{Detailed Profiling Results of DNNs Using the Profiler Module of DNNScaler for Some Representative Jobs (The higher improvement is shown by bold blue)} 
\label{tab:profIma}
\scriptsize
\setlength\tabcolsep{2pt} \begin{tabular}{lccccc}
\hline
\multirow{2}{*}{Job \#} & \multirow{2}{*}{\begin{tabular}[c]{@{}c@{}}Base Throughput\\ (BS=1 \& MTL=1)\end{tabular}} & \multicolumn{2}{c}{Multi-Tenancy Throughput}                                   & \multicolumn{2}{c}{Batching Throughput}                                       \\ \cline{3-6} 
                        &                                                                                           & MTL=8   & \begin{tabular}[c]{@{}c@{}}Improvement (\%)\\  (TI\_MT)\end{tabular} & BS=32   & \begin{tabular}[c]{@{}c@{}}Improvement (\%)\\  (TI\_B)\end{tabular} \\ \hline
1                       & 118.66                                                                                    & 237.28  & \textcolor{blue}{\textbf{99.96}}                                                                & 125.67  & 5.91                                                                \\
2                       & 104.46                                                                                    & 169.85  & \textcolor{blue}{\textbf{62.59}}                                                                & 125.33  & 19.97                                                               \\
3                       & 36.81                                                                                     & 39.61   & 7.63                                                                 & 116.41  & \textcolor{blue}{\textbf{216.28}}                                                              \\
9                       & 48.49                                                                                     & 148.28  & \textcolor{blue}{\textbf{205.81}}                                                               & 125.44  & 158.70                                                              \\
10                      & 103.62                                                                                    & 137.43  & \textcolor{blue}{\textbf{32.63}}                                                                & 126.55  & 22.13                                                               \\
11                      & 62.75                                                                                     & 78.63   & 25.32                                                                & 125.99  & \textcolor{blue}{\textbf{100.79}}                                                              \\
15                      & 102.82                                                                                    & 169.31  & 64.67                                                                & 235.05  & \textcolor{blue}{\textbf{128.61}}                                                              \\
19                      & 241.14                                                                                    & 1050.58 & \textcolor{blue}{\textbf{335.67}}                                                               & 267.84  & 11.07                                                               \\
26                      & 492.00                                                                                    & 2163.80 & 339.80                                                               & 7145.89 & \textcolor{blue}{\textbf{1352.43}}                                                             \\
29                      & 15.46                                                                                     & 41.27   & \textcolor{blue}{\textbf{166.89}}                                                               & 19.82   & 28.16                                                               \\ \hline
\end{tabular}
\end{table}

\subsection{Throughput and Power Efficiency}\label{subsec:thro}

\begin{figure*}[]
\centering
\includegraphics[width=.87\linewidth]{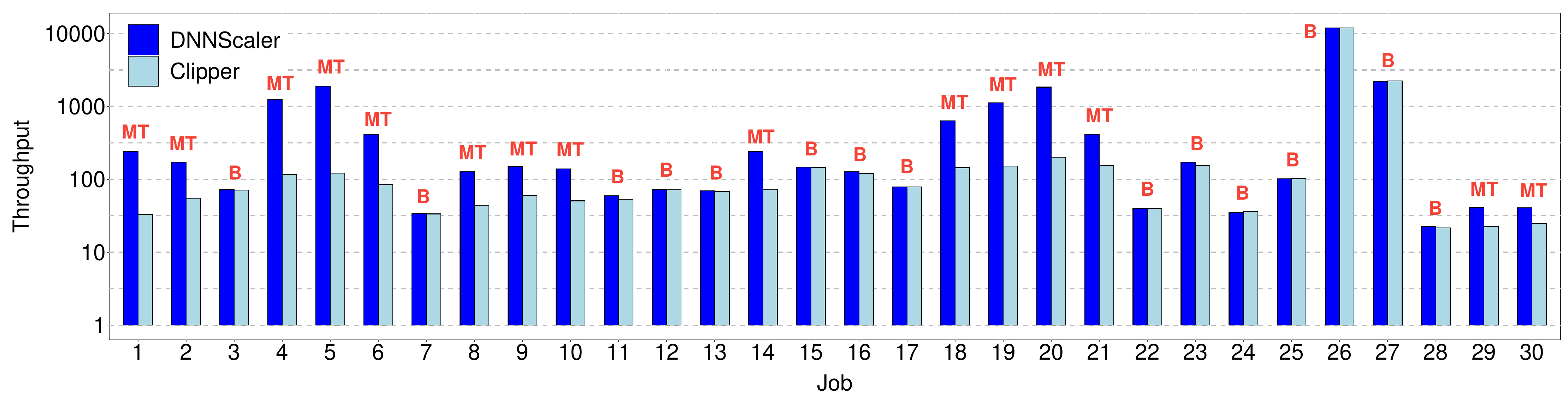}
\vspace{-10pt}
\caption{Comparing the throughput of DNNScaler and Clipper for all the jobs. The Y axis is shown in base-10 log scale. The B (Batching) and MT (Multi-Tenancy) on top of the bars indicate the selected approach by DNNScaler for that job.}\label{fig:averageThroughput}
\vspace{-5pt}
\end{figure*}

Throughput is a crucial parameter factored in when designing and deploying real-time ML services \cite{gupta2020architectural, chung2018serving}. Therefore we study the throughput of \textit{DNNScaler} to understand how much it can improve the performance of applications compared with Clipper. Fig. \ref{fig:averageThroughput} shows the throughput of \textit{DNNScaler} and Clipper for all the jobs. Note that a base-10 log scale is used for the Y axis. On average, \textit{DNNScaler} improves the throughput by 218\% compared with Clipper. For the jobs that \textit{DNNScaler} uses the Batching approach, the improvement is not as very significant (e.g., 1\% improvement in Job 7). However, for jobs such as Jobs 1 and 2 where \textit{DNNScaler} employs the Multi-Tenancy approach, the throughput improvement is as significant as 14x (Job 5). We clearly see that our proposed Multi-Tenancy approach, which determines the number of co-located instances dynamically with respect to SLO, can successfully leverage the GPU resources and significantly increase the throughput, compared with Batching strategy of Clipper. These results confirm our earlier observation that wise usage of Multi-Tenancy for some DNNs can better utilize the GPU resources, and hence, yield better throughput than Batching.

Another essential feature of real-time ML systems is power efficiency. Power is of high importance in warehouse-scale infrastructures and datacenters since it has a substantial share in operational costs \cite{goudarzi2015hierarchical, manousakis2015coolprovision}. We compare \textit{DNNScaler} and Clipper regarding power efficiency as well. We define power efficiency as the throughput per watt achieved by each approach. Since there is not much difference between performance of \textit{DNNScaler} and Clipper, when \textit{DNNScaler} uses Batching, only the results for jobs performed using Multi-tenancy by \textit{DNNScaler} are shown in Table \ref{tab:power}. 

Clipper employing large batches leads to high power consumption, but without expected throughput improvement, which leads to poor power efficiency. \textit{DNNScaler}, on the other hand, can achieve high throughput using Multi-Tenancy, and hence, better power efficiency. While \textit{DNNScaler} consumes more power than Clipper (44\% on average for jobs shown in Table \ref{tab:power}), its throughput improvement (435\% on average) can definitely compensate for it. Hence, \textit{DNNScaler} can deliver significant power efficiency improvement compared with Clipper (up to 11x in Job 5, and 288\% on average). We can conclude that \textit{DNNScaler} can remarkably enhance the power efficiency of real-time ML infrastructures.
 
\begin{table}[]
\centering
\caption{Comparing the Power Efficiency and Throughput of DNNScaler and Clipper}
\label{tab:power}
\setlength\tabcolsep{3pt} \scriptsize
\begin{tabular}{lcccccc}
\hline
\multirow{2}{*}{Job \#} & \multicolumn{2}{c}{Power (W)} & \multicolumn{2}{c}{Throughput} & \multicolumn{2}{c}{\begin{tabular}[c]{@{}c@{}}Power Efficiency \\ (Throughput/Power)\end{tabular}} \\ \cline{2-7} 
                        & DNNScaler      & Clipper      & DNNScaler       & Clipper      & DNNScaler                                         & Clipper                                        \\ \hline
1                       & 87.70          & 55.04        & 241.62          & 32.88        & 2.75                                              & 0.60                                           \\
2                       & 89.82          & 57.98        & 172.26          & 54.81        & 1.92                                              & 0.95                                           \\
4                       & 74.96          & 54.61        & 1254.10         & 116.08       & 16.73                                             & 2.13                                           \\
5                       & 63.04          & 51.78        & 1888.50         & 121.57       & 29.96                                             & 2.35                                           \\
6                       & 90.58          & 59.96        & 415.70          & 84.59        & 4.59                                              & 1.41                                           \\
8                       & 71.57          & 55.74        & 127.60          & 44.02        & 1.78                                              & 0.79                                           \\
9                       & 73.33          & 57.88        & 150.60          & 60.54        & 2.05                                              & 1.05                                           \\
10                      & 118.06         & 64.17        & 138.84          & 50.63        & 1.18                                              & 0.79                                           \\
14                      & 87.74          & 57.32        & 239.30          & 71.89        & 2.73                                              & 1.25                                           \\
18                      & 109.84         & 65.80        & 634.90          & 144.58       & 5.78                                              & 2.20                                           \\
19                      & 75.94          & 54.34        & 1118.60         & 151.41       & 14.73                                             & 2.79                                           \\
20                      & 63.30          & 52.41        & 1839.80         & 200.78       & 29.07                                             & 3.83                                           \\
21                      & 90.63          & 65.25        & 414.50          & 155.09       & 4.57                                              & 2.38                                           \\
29                      & 122.44         & 86.39        & 40.93           & 22.51        & 0.33                                              & 0.26                                           \\
30                      & 132.19         & 88.98        & 40.72           & 24.72        & 0.31                                              & 0.28                                           \\ \hline
\end{tabular}
\end{table}

\subsection{DNNScaler Results in Detail}\label{subsec:deta}
\begin{figure}[t]
\centering
\includegraphics[width=.92\linewidth]{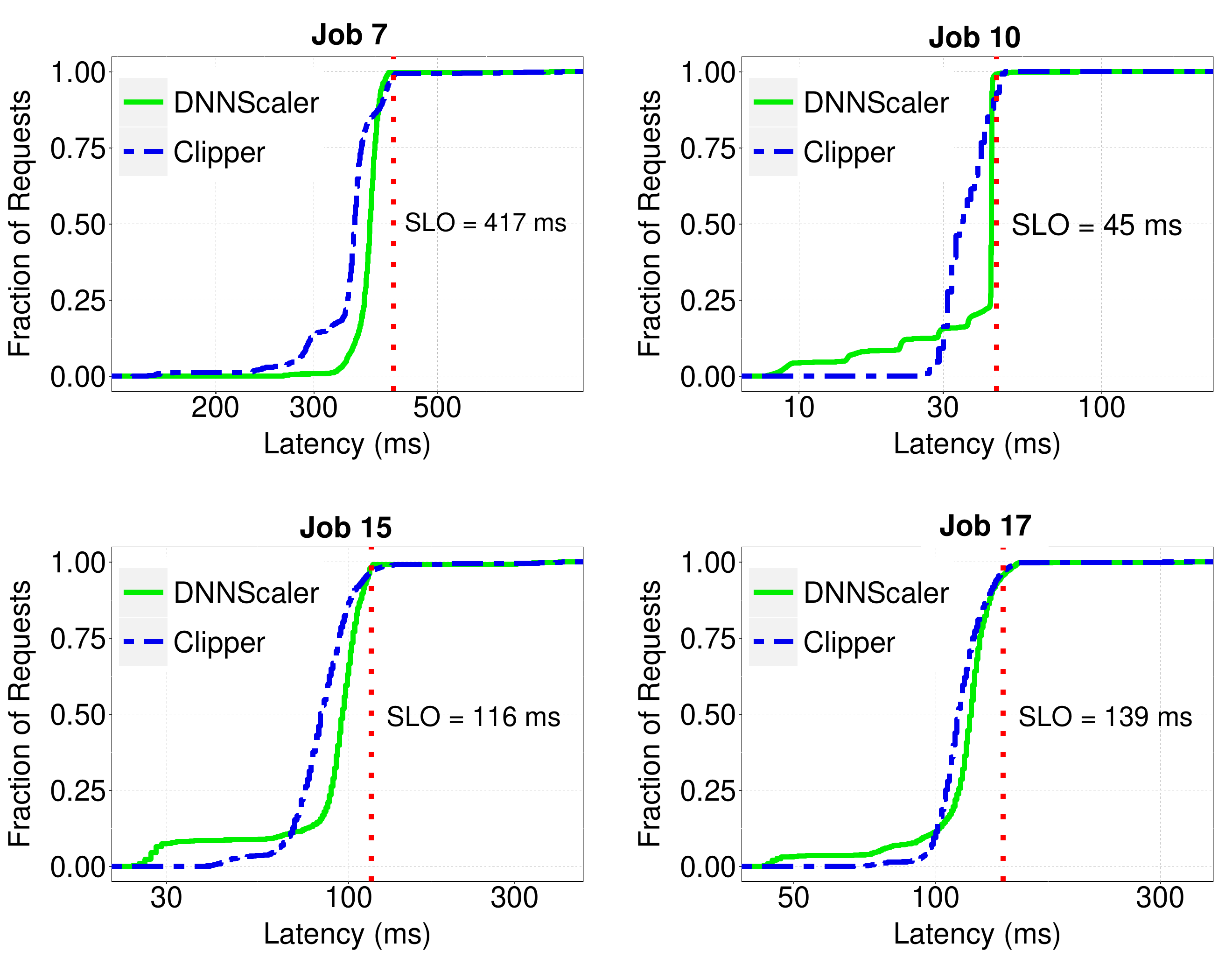}
\vspace{-10pt}
\caption{Cumulative distribution of the latency of requests for four jobs. The vertical dotted red line shows the SLO of each job. The X axis is shown in base-10 log scale.}\label{fig:cumu}
\vspace{-5pt}
\end{figure}

To better understand the behavior of \textit{DNNScaler}, we go deeper into the details and discuss the results for a few jobs. First, we depict the latency trace of a few jobs to show how both \textit{DNNScaler} and Clipper can meet the SLO. The cumulative distribution of latency of requests for four jobs is depicted in Fig. \ref{fig:cumu}. As can be seen, for both \textit{DNNScaler} and Clipper, 95\% or more of the requests have a latency smaller or equal to SLO. This emphasizes the success of both approaches in meeting the SLO of jobs.

\begin{figure}[t]
\centering
\includegraphics[width=.96\linewidth]{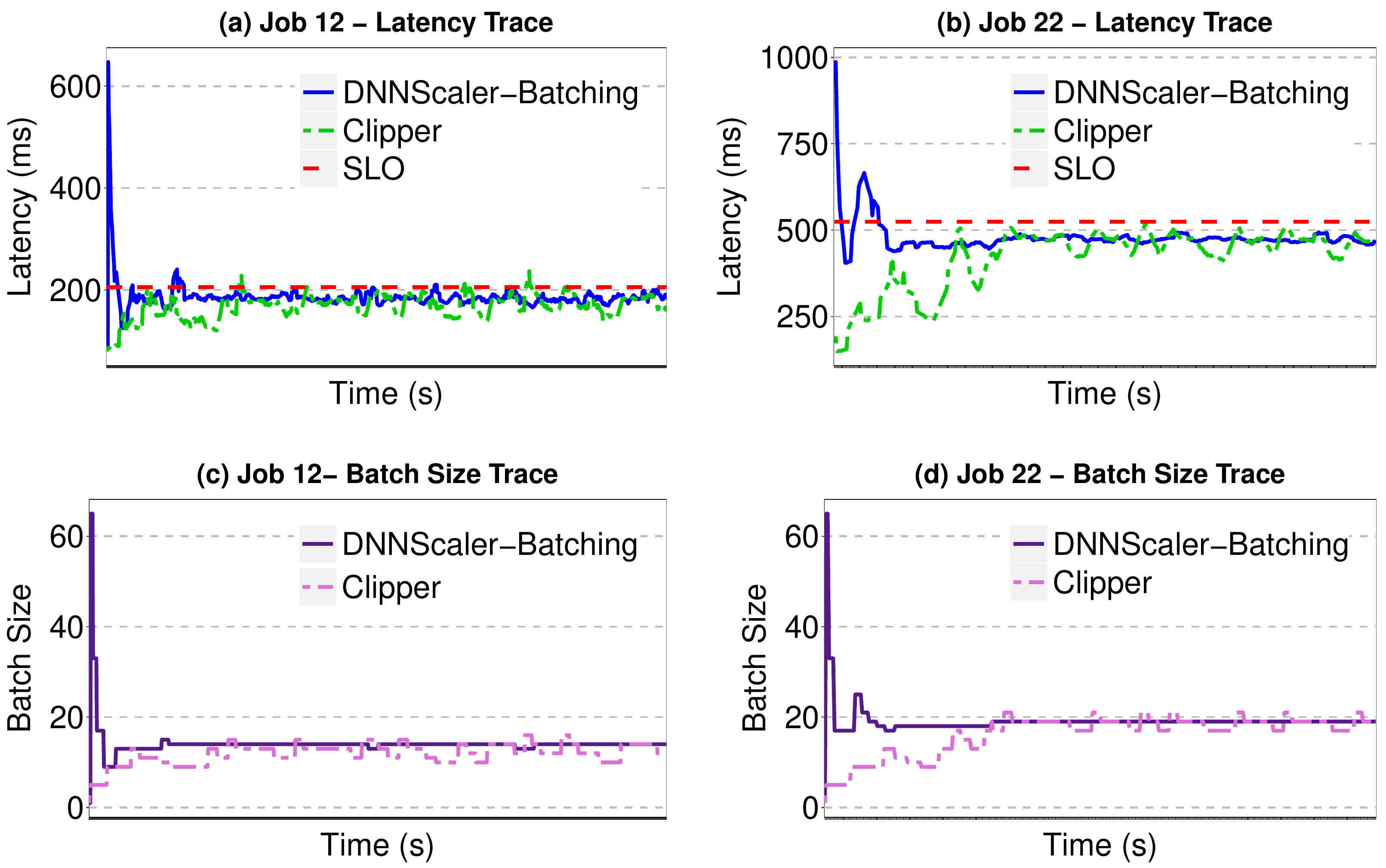}
\vspace{-10pt}
\caption{Detailed behavior of DNNScaler (Batching) and Clipper for two representative jobs.} \label{fig:detailBatching}
\end{figure}

Next, we discuss the Batching results for two jobs in Fig. \ref{fig:detailBatching}(a)(c) and Fig. \ref{fig:detailBatching}(b)(d), respectively. Since Clipper also uses the Batching mechanism, we show its results as well and compare them with \textit{DNNScaler}. Both \textit{DNNScaler} and Clipper start with BS = 1. Since the initial latency is lower than SLO, both of them increase batch size to achieve higher throughput. The pseudo-binary mechanism of \textit{DNNScaler} jumps to a relatively big batch size, but when it detects the significant SLO violation, it immediately reduces the batch size and finds the suitable one after trying a few ones by applying its search routine. The Clipper's AIMD mechanism tries to adjust the batch size as well, but with a slower rate than \textit{DNNScaler}, and consequently, it reaches the stable state later than \textit{DNNScaler}. The ability of \textit{DNNScaler} to quickly find the suitable batch size helps it to immediately adapt to possible changes in SLO or slowdown of GPU due to an applied power cap. Later in Section \ref{subsec:sens}, we study the behavior of \textit{DNNScaler} under varying SLO.

\begin{figure}[t]
\centering
\includegraphics[width=.96\linewidth]{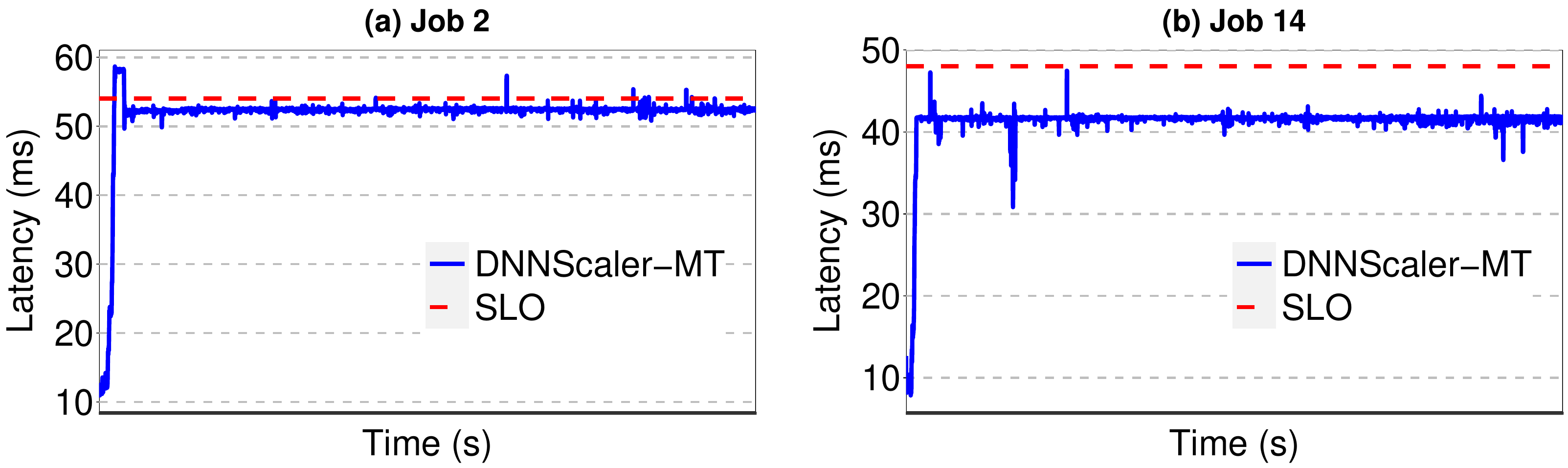}
\vspace{-10pt}
\caption{Detailed behavior of DNNScaler under Multi-Tenancy approach for two representative jobs.}\label{fig:detailMT}
\end{figure}

After that, we explore the \textit{DNNScaler} behavior for the Multi-Tenancy approach. The detailed results for two jobs are depicted in Fig. \ref{fig:detailMT}. For Job 2 (Fig. \ref{fig:detailMT}(a)), \textit{DNNScaler} initially employs the latency estimations of different MTLs from matrix completion and compares them with the SLO to decide the maximum number of DNN instances it should launch to maximize the throughput, while meeting the latency constraint. But it detects the SLO violation after launching the instances, meaning that the estimation was not 100\% accurate (as expected). Hence, it terminates one instance and since the new latency meets the SLO, it continues with the remaining number of instances. For Job 14, \textit{DNNScaler} deploys the maximum number of allowed instances (MTL = 10) based on matrix completion estimation. After deployment of the instances, the latency is still below SLO, but since there is no room for adding extra instances on GPU, it continues with current ones. 

In both Batching and Multi-Tenancy, some short-live spikes are observed in latency that violate the SLO. They happen due to some reasons (e.g., OS processes) rather than DNNScaler settings. Thereby, they are skipped to avoid excessive changes which leads to performance degradation. For long spikes that affect the tail latency significantly, DNNScaler readjusts the control knobs to mitigate them according to Algorithm 1.

\subsection{Sensitivity Analysis}\label{subsec:sens}

\begin{figure}
\centering
\includegraphics[width=\linewidth]{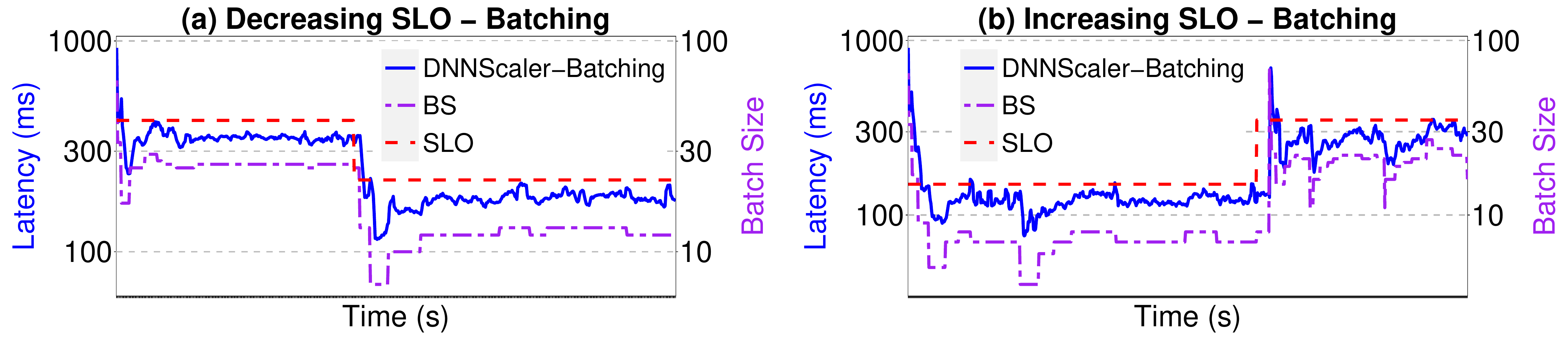}
\vspace{-10pt}
\caption{Sensitivity analysis results for DNNScaler under Batching approach for Inception-V4 network.}\label{fig:sensBatching}
\vspace{-5pt}
\end{figure}

\begin{figure}
\centering
\includegraphics[width=\linewidth]{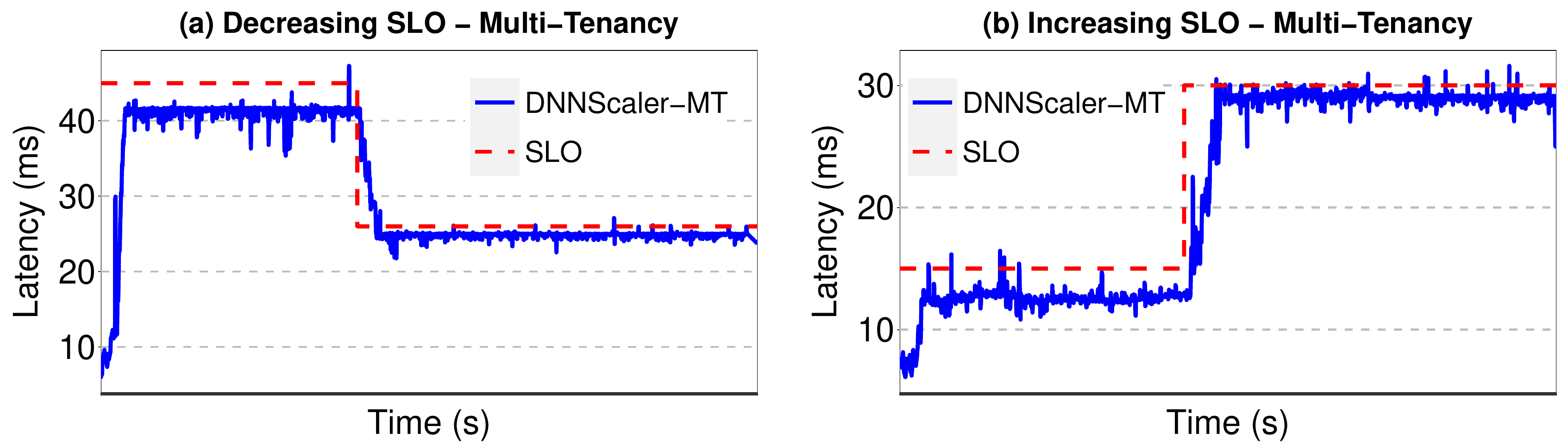}
\vspace{-10pt}
\caption{Sensitivity analysis results for DNNScaler under the Multi-Tenancy approach for Inception-V1 network.} \label{fig:sensMT}
\vspace{-5pt}
\end{figure}

During the execution of an application, any change in external and/or internal parameters of the system can affect the latency and/or the SLO of the applications. For example, the user that has submitted the job might decide to change the SLO, or a power cap might be applied on GPU, that affects its frequency, and consequently, the latency of the job. As we have mentioned earlier in Section \ref{subsubsec:scaler}, \textit{DNNScaler} is designed to adapt to such changes during the runtime. To evaluate the efficacy of \textit{DNNScaler} to adapt to such changes, we conduct another set of experiments. We consider two scenarios for each approach of \textit{DNNScaler} (Multi-tenancy and Batching). In one scenario, the SLO decreases during the runtime of the job, and in the other one, it increases. For Multi-Tenancy, we employ the Inception-V1, and for Batching we employ Inception-V4. The results are shown in Fig. \ref{fig:sensBatching} for Batching and in Fig. \ref{fig:sensMT} for Multi-Tenancy.

For the Batching approach, we see how the \textit{DNNScaler} adaptively changes the batch size to meet the new SLO. In Fig \ref{fig:sensBatching}(a), as the SLO drops down, \textit{DNNScaler} significantly reduces the batch size to avoid SLO violation. On the other hand, in the presence of an increasing SLO (see Fig. \ref{fig:sensBatching}(b)), \textit{DNNScaler} tries to employ a larger batch size to gain higher throughput. Note that a base-10 log scale is used for the Y axis. The left Y axis shows the latency and the right one shows the batch size.

As can be seen in Fig. \ref{fig:sensMT}(a), when the SLO is relaxed, \textit{DNNScaler} creates ten instances of the DNN model and deploys them on the GPU to increase the throughput. As the SLO is reduced by almost half, \textit{DNNScaler} immediately detects it and starts terminating the extra instances to meet the SLO. It eliminates five instances to meet the new tight SLO. In Fig \ref{fig:sensMT}(b), we see the results for \textit{DNNScaler} (Multi-Tenancy approach) in an increasing SLO scenario. At first, \textit{DNNScaler} only creates four instances for deployment on GPU. In the middle of the execution, when it detects new increased SLO, it deploys more instances to exploit the gap between SLO and latency in favor of throughput. 

\subsection{Discussion}\label{subsec:diss}
\textbf{Sole Employment of Multi-Tenancy.}
No previous work has used the Multi-Tenancy approach to increase the throughput of a single DNN, similar to our work. Therefore, in the experimental results section, we have no comparison with an approach that only uses Multi-Tenancy for all the jobs (i.e., we have Clipper that only uses Batching for all the jobs, but there is not an approach that only uses Multi-Tenancy). Not having such comparison makes it difficult to understand what is the impact of Multi-Tenancy on jobs such as 3 and 7 (see Table \ref{tab:jobs}) and how the Batching approach selected by \textit{DNNScaler} can improve their throughput compared against Multi-Tenancy.

\begin{figure}
\centering
\includegraphics[width=.9\linewidth]{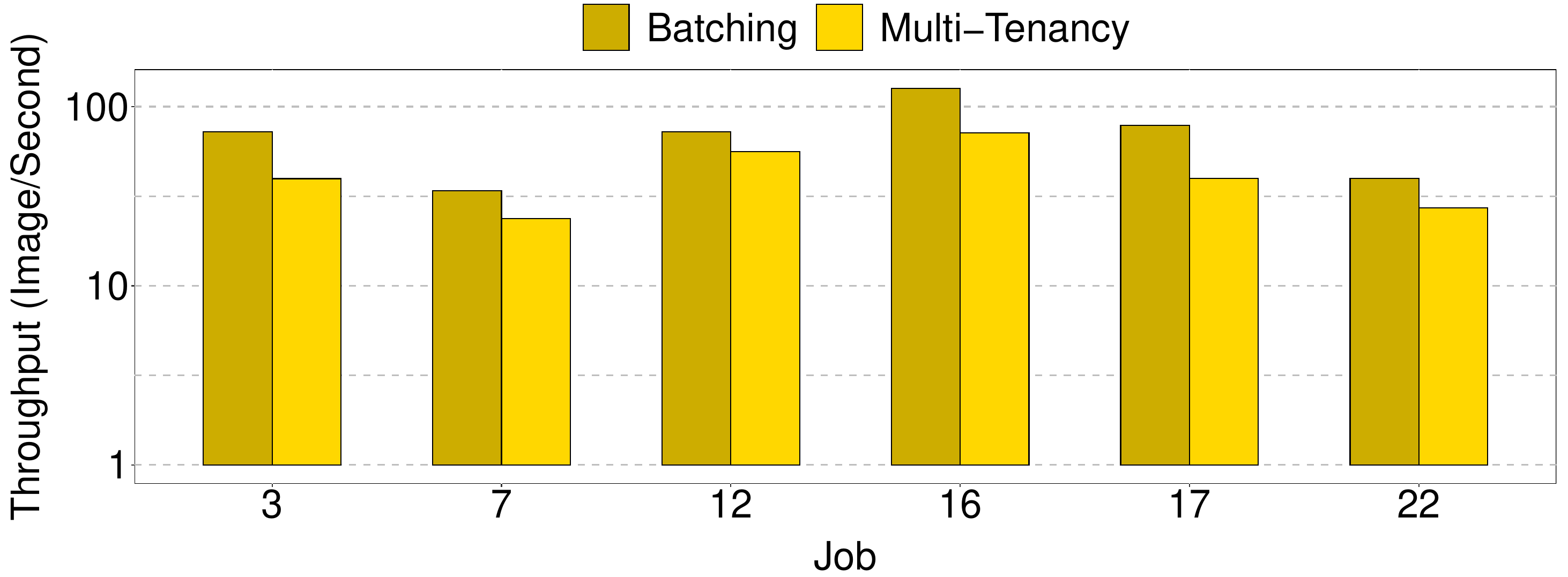}
\vspace{-10pt}
\caption{Comparing the throughput under Batching and Multi-Tenancy. DNNScaler selects Batching for these jobs.} \label{fig:discuss}
\vspace{-5pt}
\end{figure}

To address these questions, in this section, we compare the throughput of the Batching and Multi-Tenancy approaches for 6 jobs from Table \ref{tab:jobs} that have been executed by the Batching approach (according to \textit{DNNScaler} decision) in earlier experiments. The Multi-Tenancy approach we have used for these jobs is exactly the Multi-Tenancy approach of \textit{DNNScaler} that has been used for other jobs. The result is shown in Fig. \ref{fig:discuss}. \textit{Our goal is to verify that the \textit{DNNScaler's} decision of employing the Batching approach for these jobs was a correct one, and that Multi-Tenancy cannot improve their throughput more than Batching.} We see that in all the jobs, Batching yields higher throughput than Multi-Tenancy, so we conclude that \textit{DNNScaler} has selected the correct approach for them.

\noindent\textbf{Combining Batching and Multi-Tenancy}. A question that can rise is that why not combining the two approaches to leverage benefits of both of them. For example, in the Multi-Tenancy we mentioned that we use BS = 1 for all the instances. But, what if we use larger batch size for them? To answer this question, we conduct another set of experiments. We consider two DNNs that were executed by Batching (ResV2-152 and PNAS-Large) and two ones that were executed by Multi-tenancy (MobV1-1 and MobV1-025). For ResV2-152 and PNAS-Large, we select a fixed batch size of 8 (BS=8, constant) and increase the number of co-located instances from 1 to 4 (MTL=1 to MTL=4), and measure the throughput and latency of the DNNs for each MTL. ResV2-152 experiences notable throughput improvement when going from MTL=1 to MTL=2, however, the improvement for MTL=3 and MTL=4 is negligible. PNAS-Large, on the other hand, not only experiences no throughput improvement, but even suffers from reduction in throughput. As expected, the latency of both of them increases as enlarging the MTL.

For MobV1-1 and MobV1-025, we consider a fixed number of co-located instances of 5 (MTL=5, constant) and change the batch size from 1 to 8 (BS=1, BS=2, BS=4, BS=8). Again, we see that one of them (MobV1-1) can benefit from the combination of Batching and Multi-Tenancy in term of throughput, while the other one (MobV1-025) experiences no throughput improvement and only suffers from higher latency. We observe that the largest network (PNAS-Large) and the smallest one (MobV1-025) cannot benefit from the combination, while the two other ones can benefit up to a certain level. We conclude that combining the Batching and Multi-Tenancy can lead to throughput improvement in some DNNs, but for the other ones it only elongates the inference latency with no benefits in throughput. Hence, identifying the proper cases of combining Batching and Multi-Tenancy based on different aspects of the system such as the size and computational complexity of DNNs, as well as the computing and memory capacity of the GPU, can be a future research direction.

\begin{figure}
\centering
\includegraphics[width=\linewidth]{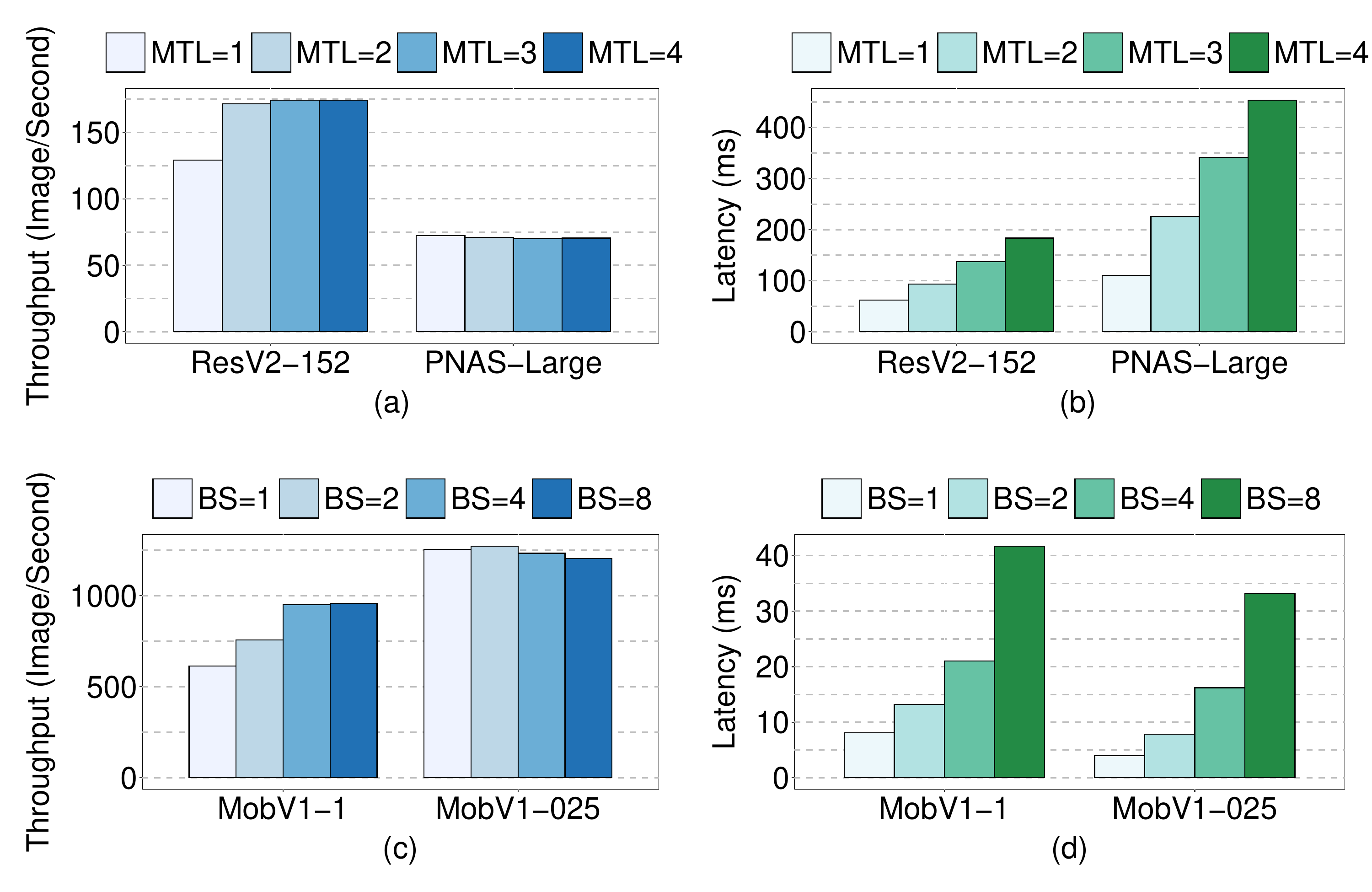}
\vspace{-10pt}
\caption{Studying the impact of Batching and Multi-Tenancy combination on the performance of DNNs.}
\label{fig:combin}
\vspace{-5pt}
\end{figure}

\section {Related Work}\label{sec:rela}

While DNNs continue to deliver state-of-the-art results for various machine learning domains such as computer vision, their extremely growing computational requirements have surpassed the growth in computing capacity of conventional CPUs \cite{chung2018serving}. Therefore, it is essential to investigate new hardware platforms, beyond traditional CPUs, to address the ever-growing computational demand of DNNs. To this end, a wide variety of DNN accelerators are designed and implemented that aim to achieve various goals such as low-latency, high-throughput, or energy-efficiency \cite{zhu2016lradnn, shafiee2016isaac, chen2014diannao, gao2018low, chung2018serving, jiang2019achieving}. Increasing throughput while achieving low-latency is specially explored to address the requirements of real-time ML services deployed on warehouse-scale infrastructures \cite{fowers2018configurable, gupta2020architectural, gupta2020deeprecsys}. These accelerators employ different computing cores such as ASICs \cite{chen2016eyeriss, jouppi2017datacenter, chen2014dadiannao}, FPGAs \cite{zhang2018dnnbuilder, wei2019overcoming, li2016high}, and GPUs \cite{shen2019nexus, wang2020sparsert, li2020automating} or different computing paradigms such as processing in/near-memory \cite{ueyoshi2018quest, gao2017tetris, ankit2019puma, ji2019fpsa, chi2016prime} for accelerating DNN inference. GPU accelerators are a favorable choice for DNN accelerators due to their programmability and scalability features \cite{nabavinejad2020overview}. To further improve the performance of DNN inference on GPU accelerators, various techniques such as Batching and Multi-Tenancy are proposed, among others.

\noindent\textbf{Batching.} Using Batching to increase the DNN inference throughput has been studied and employed in a large body of previous works \cite{fang2017qos,7551407, tang2019nanily, shen2017escher, zhang2020dybatch, song2017towards}. Studies show that Batching can improve the throughput and energy-efficiency of DNN inference on GPU accelerators \cite{fang2017qos, inoue2019queueing}. However, it elongates the latency of DNN inference as well, so it should be employed carefully. Pervasive CNN (P-CNN) \cite{song2017towards} leverages Batching to improve the throughput of CNNs on GPUs. It uses big batch sizes for background tasks to maximize throughput and reach energy-efficiency. When selecting the batch size for such tasks, P-CNN considers the GPU memory. For interactive and real-time tasks, however, P-CNN selects small batch sizes to avoid unacceptable response time. Clipper \cite{crankshaw2017clipper} forms batches of inputs from concurrent stream of prediction queries to leverage the benefits of Batching. It dynamically changes the batch size using an additive-increase-multiplicative-decrease (AIMD) scheme to find the optimal one that maximizes the throughput while meeting the latency requirement.    

\noindent\textbf{Multi-Tenancy.} A large body of research has focused on challenges and opportunities of Multi-Tenancy and co-location of DNN inference \cite{wei2020predicting, jeon2019analysis, wang2017quality, choi2020prema}. PERSEUS \cite{lemay2020perseus} and Jain et al. \cite{jain2018dynamic} studied the impact of Multi-Tenancy on performance, cost, and latency of co-located DNN models. They showed that while co-location can help to improve the throughput, resource utilization, and energy efficiency of GPUs, it has a negative impact on latency. Approaches such as Baymax \cite{chen2016baymax} and Laius \cite{zhang2019laius} try to mitigate the impact of co-location on the latency of interactive jobs that share the GPU with throughput-oriented jobs. They aim to maximize the throughput of throughput-oriented job while meeting the latency of interactive job by reallocation of time slots \cite{chen2016baymax} or computing resources \cite{zhang2019laius} of GPU. These approaches usually consider the co-location of two or more different applications on a GPU and try to manage their latency or throughput with respect to some priority criteria. In our approach, however, we consider the case where a various number of instances from the same application are co-located on a GPU, and we try to improve the overall throughput of that application while meeting its latency SLO. Moreover, we first evaluate that application to see if this type of Multi-Tenancy can help to improve its throughput, and then proceed with the next steps. But the other approaches usually consider the throughput of the mixture of applications, and not a single one.

\section{Conclusion}\label{sec:conc}

In this paper, we performed an extensive set of analysis, revealing that DNNs can be categorized in two groups: the ones that experience high throughput from Batching and the ones that achieve high throughput by Multi-Tenancy. Based on this observation, we proposed the \textit{DNNScaler} approach to improve the throughput of real-time ML services with latency constraint. The \textit{DNNScaler} Profiler module can successfully determine the approach that is more suitable for a specific DNN with a lightweight profiling mechanism. Based on the output of the Profiler module, the Scaler module employs one of the adaptive batching (for the Batching approach) or instance co-location management (for the Multi-Tenancy approach) to maintain the latency while maximizing the throughput. The experimental results show that DNNScaler can improve the throughput by up to 14x (218\% on average) compared to the Clipper approach that only leverages Batching, and not Multi-Tenancy. Furthermore, we analyzed the sensitivity of both Batching and Multi-Tenancy approaches of \textit{DNNScaler}.

\bibliographystyle{plain}
\bibliography{refs}

\end{document}